\documentclass[letterpaper,twocolumn,10pt]{article}
\usepackage{usenix-2020-09}

\usepackage{tikz}
\usepackage{amsmath}

\usepackage{enumitem}
\usepackage{pifont}
\usepackage{wasysym}
\usepackage{cleveref}
\usepackage{makecell}

\setitemize{noitemsep,topsep=0pt,parsep=0pt,partopsep=0pt}
\usepackage{microtype}
\usepackage{dblfloatfix}
\usepackage{url}

\microtypecontext{spacing=nonfrench}

\newcommand{\bptree}{{B$^+$-tree}}

\begin{document}

\date{}

\title{\Large \bf Employ SmartNICs' Data Path Accelerators for Ordered Key-Value Stores}

\author{
{\rm Frederic Schimmelpfennig}\\
Johannes Gutenberg \\University Mainz
\and
{\rm Jan Sass}\\
Johannes Gutenberg \\University Mainz
\and
{\rm Reza Salkhordeh}\\
Johannes Gutenberg \\University Mainz
\and
{\rm Martin Kröning}\\
RWTH Aachen \\University
\and
{\rm Stefan Lankes}\\
RWTH Aachen \\University
\and
{\rm André Brinkmann}\\
Johannes Gutenberg \\University Mainz
} 

\maketitle

\begin{abstract}

Remote in-memory key-value (KV) stores serve as a cornerstone for diverse modern workloads, and high-speed range scans are frequently a requirement. However, current architectures rarely achieve a simultaneous balance of peak efficiency, architectural simplicity, and native support for ordered operations. Conventional host-centric frameworks are restricted by kernel-space network stacks and internal bus latencies. While hash-based alternatives that utilize OS-bypass or run natively on SmartNICs offer high throughput, they lack the data structures necessary for range queries. Distributed RDMA-based systems provide performance and range functionality but often depend on stateful clients, which introduces complexity in scaling and error handling. Alternatively, SmartNIC implementations that traverse trees located in host memory are hampered by high DMA round-trip latencies.

This paper introduces a KV store that leverages the on-path Data Path Accelerators (DPAs) of the BlueField-3 SmartNIC to eliminate operating system overhead while facilitating stateless clients and range operations. These DPAs ingest network requests directly from NIC buffers to navigate a lock-free learned index residing in the accelerator's local memory. By deferring value retrieval from the host-side tree replica until the leaf level is reached, the design minimizes PCIe crossings. Write operations are staged in DPA memory and migrated in batches to the host, where structural maintenance is performed before being transactionally stitched back to the SmartNIC. Coupled with a NIC-resident read cache, the system achieves 33 million operations per second (MOPS) for point lookups and 13 MOPS for range queries. Our analysis demonstrates that this architecture matches or exceeds the performance of contemporary state-of-the-art solutions, while we identify hardware refinements that could further accelerate performance.
\end{abstract}


\section{Introduction}

Many modern workloads rely on remote key-value (KV) stores that support lookups, inserts, and range queries. However, remote performance is often limited by transport bottlenecks. Traditional remote KV stores such as \textsc{Redis}~\cite{redis} and \textsc{Memcached}~\cite{memcached} are accessed via Ethernet. Their use of the operating system (OS) kernel network stack and PCIe bottlenecks between the network interface card (NIC) and the host limit throughput and increase latencies.

To mitigate OS overhead, some approaches use the Data Plane Development Kit (DPDK)~\cite{dpdk} to bypass the network stack via direct user-space access. \textsc{Mica} reaches nearly 100 million operations per second (MOPS) on a single server node~\cite{LimHAK14}. Other works like \textsc{KV-Direct} achieve even higher throughput by removing the host involvement and serving requests directly from an FPGA SmartNIC~\cite{LiRXLXPCZ17}. This reduces latencies because the KV store is closer to the network. However, to achieve these results, \textsc{KV-Direct} is constrained in the complexity and capacity of the underlying data structures. Both \textsc{Mica} and \textsc{KV-Direct} use hash-based point lookups and lack range queries. Recent works apply similar offload designs to more recent SmartNIC generations and are affected by the same shortcomings~\cite{KashyapLL25,PismennyL0T22}. Approaches like \textsc{Honeycomb}~\cite{LiuDFKKZNKC24} traverse data structures in host memory and support range queries. However, their performance falls behind because they issue frequent DMA round-trips.

Clients of KV stores that use remote direct memory access (RDMA) can  access host memory directly~\cite{ZVBFK19, LiHZCS23, Li0ZCS23, WeiCCZ21, WangW0S25, RenZCXCW24, Wang0KOA23}.
These approaches introduce architectural challenges because every client eventually needs to know the remote addresses of the data on the server to access it. \textsc{Sherman} minimizes round trips for writes but requires client caches for tree traversals~\cite{LiHZCS23}. Similarly, \textsc{ROLEX}~\cite{Li0ZCS23} maintains metadata indexes on the clients and tries to optimize accesses using learned indexes.
These systems can deliver high performance but depend on client-side logic and resources, which raises scaling complexities, introduces consistency concerns, and requires additional failure handling.

In this paper, we present \emph{DPA-Store}, an ordered in-memory KV store that uses the BlueField-3 SmartNIC~\cite{nvidia-bluefield} and its \emph{Data Path Accelerators} (DPAs)~\cite{bluefield3-dpa} to bypass OS overheads and to allow stateless clients. DPAs are highly parallel, programmable compute units embedded in the network data path. DPAs have direct access to NIC buffers and NIC-side DRAM and can perform DMA operations to access host memory, enabling low-latency, flexible request processing~\cite{Chen0FSMQZSZLW24}. 

DPA-Store is built around the high level of concurrency of the DPA subsystem, which consists of 16 cores, each with 16 threads, for a total of 256 threads. When receiving requests, these threads traverse a learned index tree~\cite{KraskaBCDP18} stored in NIC memory without host involvement. To circumvent memory capacity restrictions of the NIC, values are stored in a replica of the tree on the host. Both point and range lookups require the DPAs to perform a minimal number of DMA operations when reaching a leaf node to access values in host memory. This minimizes the number of times the PCIe boundary between the NIC and the host is crossed. 

When an insert request is received, the responsible DPA thread appends the entry to an insert buffer at the leaf level on the NIC-side. These entries are immediately visible for subsequent lookups. The insert buffers are transferred in batches to the host when they are full. The host, with its greater compute capabilities, performs expensive structural updates to its tree replica using concurrent \emph{patch} threads that apply the inserts to the host-side tree, retrain affected sub-trees, and prepare new nodes. A set of \emph{stitch} threads then copies the new nodes into the NIC-side tree and makes them available via pointer swaps. This guarantees consistency while keeping the NIC-side traversal path lock-free.

We employ a NIC-side read cache for hot entries to reduce the number of tree traversals and DMA operations. Requests are routed to DPA threads based on their hash value, so that small Bloom filters to guard cache accesses fit in the remaining cache line space of each DPA thread. These cache lines are nearly never evicted from the L1 caches of the DPA threads, allowing efficient caching of hot key-value pairs.

Our learned index allows fine-grained optimizations for the DPA subsystem’s memory characteristics, scheduling, and concurrency. We design the traversal logic around optimal cache line accesses and employ fixed-point calculations to compensate for the lack of floating-point units on the DPAs. Our contributions are as follows:

\begin{itemize}[leftmargin=1.25em,topsep=0.2ex,itemsep=0.3ex]
    \item We show that it is possible to build a KV store supporting lookups, inserts, and range queries on a BlueField-3 SmartNIC that is already competitive, for most operations, with state-of-the-art RDMA-based KV stores \emph{without} relying on stateful clients. DPA-Store achieves 33\,MOPS for \texttt{GET} and 13\,MOPS for \texttt{RANGE} operations.
    \item We execute extensive experiments to evaluate DPA-Store. In doing so, we reveal in-depth performance characteristics of the BlueField-3, exceeding prior works. These insights also explain the low \texttt{INSERT} performance of DPA-Store of only 1.7\,MOPS. 
    \item We demonstrate how modest modifications to the BlueField-3 hardware, without changing the overall programming model, could unlock substantial performance gains, which would easily overcome the performance of today's fastest RDMA-based KV stores.
\end{itemize}

\noindent
The paper is structured as follows. Section~2 provides background and surveys related work. Section~3 details the DPA-Store architecture and implementation. Section~4 evaluates DPA-Store with a sensitivity analysis and comparison benchmarks. Finally, Section~5 concludes the paper.

\section{Background \& Related Work}
\label{sec:background}

In this section, we summarize the background on learned indexes and the BlueField-3 DPU and position DPA-Store relative to kernel-bypass, NIC-assisted, and RDMA-based remote KV systems.

\subsection{Classical \& Learned Index Structures}

\begin{figure}[b]
\includegraphics[width=\linewidth,clip,trim={0 0 1.85cm 0}]{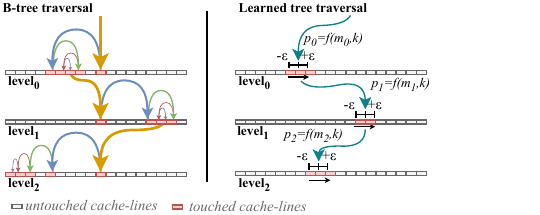}
\caption{Access patterns in B-trees vs. learned trees.
}
\label{fig:learned:index}
\end{figure}

Key-value (KV) stores expose a compact interface, providing \texttt{GET}, \texttt{INSERT}, \texttt{UPDATE}, \texttt{DELETE}, and optionally \texttt{RANGE} operations. Typically, KV stores rely on either hash tables or, if range operations are required, ordered tree structures. Modern, non-remote in-memory KV stores utilize optimized variants of traditional data structures such as B$^+$-trees, tries, or hash maps to serve node-local processes with highly optimized traversals and concurrency control~\cite{LeisK013, LeisSK016, MaoKM12, WuNJ19}. However, traditional data structures often suffer from suboptimal memory access patterns \cite{KraskaBCDP18}.

To address this issue, \emph{learned indexes}~\cite{KraskaBCDP18, MarcusZK20, DingMYWDLZCGKLK20, FerraginaV20, WuZCCWX21, GeZSLGCCP23, LiLZDYP23, KipfMRSKK020, ZhangQYB24, LiHJZ21} capture the key distribution using lightweight models, allowing lookup operations to jump directly to the relevant region inside a node before a short scan takes place, requiring only very few cache line accesses per tree level (see \Cref{fig:learned:index}). A common choice for approximating the key distribution is a piecewise linear approximation (PLA). Short linear segments are fitted to the cumulative key distribution, enabling a linear prediction $p = a \cdot k + b$ of the target position of key $k$ using parameters $a$ and $b$ with an error of at most $\varepsilon$. 

Thus, instead of $O(\log n)$ memory accesses per node with potentially many independent cache line accesses (e.g., in a B$^+$-tree), lookup operations in a learned index are aggregated in a small contiguous window $[p-\varepsilon, p+\varepsilon]$, trading fast CPU cycles for relatively slow memory accesses. 

The error bound $\varepsilon$ balances capacity efficiency against scan effort. Smaller bounds reduce the scan range but increase the number of models, whereas larger bounds keep more data in continuous segments, requiring larger regions to be scanned. Therefore, choosing a suitable parameter $\varepsilon$ is important when mapping learned indexes to accelerators like SmartNICs with their limited memory access performance.

Insert and model rebuild strategies further impact runtime behavior. \textsc{Alex}~\cite{DingMYWDLZCGKLK20}, e.g., stores keys in gapped arrays with small buffers and does not require relearning until gaps are full. \textsc{Hyper}~\cite{ZhangQYB24} combines bottom-up and top-down strategies and utilizes overflow buffers to limit memory overheads, while \textsc{XIndex}~\cite{ZhangQYB24} targets highly concurrent workloads. In all cases, buffer layout, size, and update policies govern retraining cost and concurrency characteristics.

\subsection{Remote KV Stores}
Remote KV stores introduce additional complexity by allowing access over network protocols. Widespread Ethernet-based KV stores~\cite{redis, memcached, dragonfly, keydb} provide general-purpose use. However, their request throughput is limited by the host networking stack and related operating system (OS) components, such as network sockets, context switches, and the NIC-host PCIe boundary~\cite{SchuhKCLRKS24, CaiCVH021, NeugebauerAZAL018}. 

To address these issues, RDMA-based (Remote Direct Memory Access) KV stores bypass the OS and networking stack~\cite{ZVBFK19, KaliaKA14, ChenCWSCWZG17}. They utilize traditional tree architectures~\cite{LiHZCS23, Wang0KOA23, WangW0S25} or learned indexes~\cite{WeiCCZ21, Li0ZCS23} and offer the full set of KV operations, including range queries. 

\textsc{Sherman}~\cite{LiHZCS23} is a write-optimized distributed B$^+$-tree over disaggregated memory that employs hierarchical NIC on-chip locking, client-side index caching, and lock-free reads to increase write throughput. \textsc{SMART}~\cite{RenZCXCW24} optimizes this approach by addressing RDMA-NIC scale-up bottlenecks. \textsc{\mbox{XStore}}~\cite{WeiCCZ21} couples a server-side B$^+$-tree with a learned cache on the client, which the client uses to predict the location of the KV pair on the server, saving RDMA roundtrips. This allows range queries to complete with as few as two RDMA operations. The server retrains models in the background. \textsc{ROLEX}~\cite{Li0ZCS23} addresses dynamic workloads by strictly controlling data movement within a learned tree, allowing an asynchronous retraining of models. A leaf-atomic shift scheme keeps leaves sorted and minimizes interference. 

These systems demonstrate that low-latency serving of ordered range queries can achieve high throughput. However, they require clients that execute stateful index logic or depend on multiple RDMA roundtrips per request.

Severe bottlenecks can also be overcome by minimizing the impact of the operating system or by using SmartNICs. \textsc{Mica} combines DPDK user-space networking with parallel data partitioning to bypass kernel overheads, allowing a single server to sustain nearly 100~MOPS~\cite{LimHAK14, dpdk}. \textsc{KV-Direct} offloads the KV store to FPGA-programmable NICs by extending RDMA primitives, serving NIC-resident point lookups that scale with multiple NICs per server while reaching microsecond tail latencies even at high throughput, albeit with constrained capacity~\cite{LiRXLXPCZ17}. Both \textsc{Mica} and \textsc{KV-Direct} use hash-based data structures to demonstrate unconstrained throughput and do not support range queries. 

Some recent works~\cite{KashyapLL25, PismennyL0T22, WeiCY0023, ZhangLWWLYWW23} also lack support for range operations but contribute detailed insights into current SmartNIC capabilities, which can offload increasingly complex KV-store mechanics~\cite{ThostrupFZB22, KfouryCMAGC24, SunHZCHZW23, SunZYW22, KashyapLNL25}. 

\textsc{Honeycomb}~\cite{LiuDFKKZNKC24} implements a B-tree traversal within an FPGA on a SmartNIC that accesses a tree on host memory via DMA. It provides a KV-store interface, including ranges, but requires multiple expensive DMAs for uncached KV pairs. \textsc{HiDPU}~\cite{ZhuSWCYYS25} performs address translation for disaggregated storage and uses the Huawei Hi1823 SmartNIC's DPAs to map partially continuous address areas. \textsc{HiDPU} stores those mappings in specialized segments, accessed through learned models within the 4\,MB NIC-side memory.
This work is orthogonal in terms of features and requirements to a general-purpose KV store, relies on hashed segments, and does not support range queries across multiple mappings.
Nevertheless, its performance benefits motivate offloading KV stores onto SmartNICs.
\textsc{DALdex}, on the other hand, uses SmartNICs to offload the training of a persistent memory-learned index to save host overheads, rather than for network serving~\cite{TongHC25}.

\subsection{BlueField-3 and DPA Subsystem}

\begin{figure}[t]
\includegraphics[width=\linewidth]{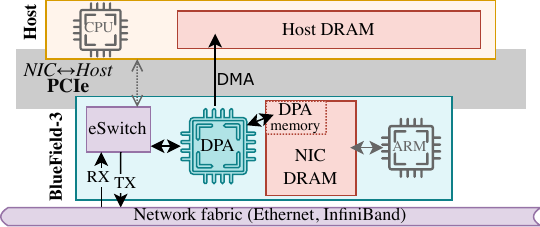}
\caption{NVIDIA BlueField-3 architecture configured in NIC mode, allowing DMA access to host memory.}
\label{fig:bf3-arch}
\end{figure}

\begin{figure*}[t]
\includegraphics[width=\linewidth]{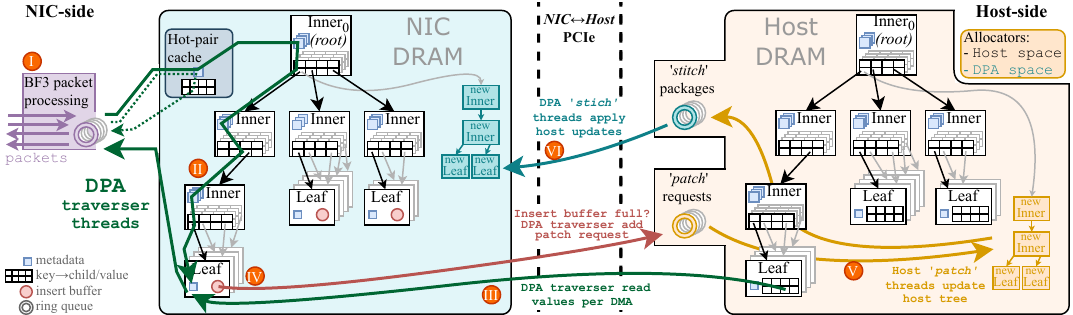}
\caption{DPA-Store architecture: (I) Request packets are assigned on-path to a DPA traverser thread, which (II) walks a NIC-side learned index. (III) Values are fetched per DMA using leaf-level models, and (IV) inserts are placed in leaf-level insert buffers. (V) The host performs structural updates that are (VI) transactionally \emph{stitched} back to the NIC without interrupting traversers.}
\label{fig:overview}
\end{figure*}

The BlueField-3 SmartNIC is a ConnectX-5-based network adapter. In addition to high-speed network functions, it offers an off-path ARM CPU and an on-path DPA cluster, as well as 16\, GIB DDR5 memory (see Figure~\ref{fig:bf3-arch}). The ARM CPU allows the BlueField-3 to run a dedicated operating system, e.g., for control plane management, and access to accelerator engines on the SmartNIC (e.g., de/encryption engines).

The DPA subsystem consists of 16 physical RISC-V cores, of which each core is running at 1.8\,GHz and features 16 threads. Currently, only 189 concurrent threads out of the 256 available threads can be used by application code~\cite{Chen0FSMQZSZLW24}. DPA threads are hardware-scheduled at fine granularity, enabling high concurrency and packet-processing pipelining.

Each DPA thread has a private L1 cache, shared 1.5\,MiB L2 and 3\,MiB L3 caches. The cache hierarchy is backed by a dedicated 1\,GiB region of the BlueField-3 DDR5 memory, termed \emph{DPA memory}. Additionally, the DPAs can access either the remainder of the BlueField-3 memory or the host memory, depending on the configuration. If the BlueField-3 is configured for DPA DMA access to the host, the ARM CPU is disabled~\cite{Chen0FSMQZSZLW24}.

DPA thread invocation can be controlled by applications on the host or the ARM or by defining a Transport Interface Receive (TIR) target for incoming network packets. For the latter, packet matching rules on hardware-defined fields (e.g., receiver or sender addresses or ports, or VLAN tags) can be used to filter relevant packets at line rate. Matched packets are put directly into the L2 cache of the DPAs, allowing low-latency access to packet data. Packets not matched are transparently placed on the default network path.

Previous works have evaluated the architecture of the BlueField-3 and the DPA subsystem~\cite{Chen0FSMQZSZLW24}. The BlueField-3 aims for a broad range of applications by featuring on-path and off-path compute, specialized accelerator engines, and dedicated memory. However, developing high-performance applications requires an in-depth understanding of the limitations of the architecture. Memory accesses to DPA memory, e.g., induce a latency of nearly 500\,ns and are significantly slower than DRAM accesses of a standard CPU. In the context of this work, it is therefore important to minimize the number of memory accesses to the tree-based data structure, motivating the use of learned indexes with their limited number of memory accesses per tree level.

\section{DPA-Store: Architecture \& Implementation}
\label{sec:architecture}

In this section, we present DPA-Store, our remote KV store that supports range queries and allows stateless clients. DPA-Store utilizes the DPA subsystem of the BlueField-3 DPU to process incoming requests, avoiding all OS overheads and leveraging the high degree of parallelism of the DPAs. 

We selected a learned index tree to reduce the number of accesses to the latency-constrained DPA memory.  Due to the restricted size of only 1\, GiB of DPA memory, DPA-Store uses the high-capacity host memory to store values in a tree replica, retrieving a value with a minimum number of DMA operations between NIC and host. Furthermore, the host is used for compute-heavy or blocking tree operations, such as model retraining and node splits. The host propagates tree updates to the NIC-side via read-copy-update (RCU) semantics, enabling the DPAs to traverse the index tree without locks and maintaining consistency at all times.

DPA-Store employs thread-local caches for hot KV pairs. Insert and update operations are batched in a leaf-level insert buffer, reducing the number of tree updates propagated to the host. Figure~\ref{fig:overview} shows the overall architecture, which is discussed in detail in this section. We start with the NIC-side request path (Sec. \ref{subsecsec:nicside-request-path}), including the learned index and caching details, and then continue with the update cycle involving the host (Sec. \ref{subsecsec:update-cycle}).

\subsection{Request Processing in the DPA Subsystem}
\label{subsecsec:nicside-request-path}

DPA-Store uses the User Datagram Protocol (UDP) over Ethernet for transporting requests, with each request consisting of a single UDP packet. All requests are terminated on the DPA threads of the BlueField-3. We have defined packet matching rules to map each of the listening UDP ports to a designated DPA thread, effectively utilizing hardware steering rules to distribute requests over available DPAs. While every request can be served from every DPA thread, the port selection can be used by clients to facilitate load balancing or to improve cache utilization. As a default, clients use the same key hashing for distributing the load uniformly across all DPA threads. 

\begin{figure}[t]
\includegraphics[width=\linewidth]{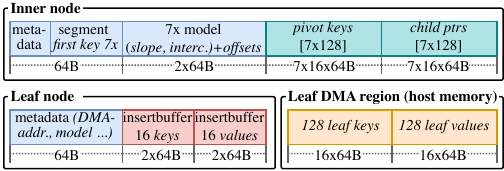}
\caption{Layout of NIC-side nodes and leaf DMA regions.}
\label{fig:memory-layout}
\end{figure}

Incoming requests are consumed by the responsible \emph{traverser} threads (176 in total), which descend the learned index tree until reaching the appropriate leaf node. 
Inner nodes contain seven learned models, whereas each model consists of a piecewise linear approximation (PLA) over its interval, pivot keys, and pointers to the respective children~(see \Cref{fig:memory-layout}). We limit the number of pivot keys and child pointers to 128 per segment, which is matched to fit our selected $\varepsilon$ values.

Partitioning inner nodes into seven segments allows us to fit the segments' first keys and node metadata into a single cache line. While the traverser thread performs a binary search, comparing the operation key with the segment keys, the segment models are prefetched, overlapping compute with memory access. Having found the target segment for a key, the segments' model is evaluated to obtain the predicted position. Again, during the computation, we prefetch the cache lines containing the pivot keys of the current segment. The prediction is used to scan the pivot array to find the index and the pointer to the designated child node before the traverser thread moves to the next level. We store pivots and child pointers separately to compare more pivots per cache line before incurring a single access to the child.

Once a traverser thread reaches a leaf node, it performs the requested operation.
\texttt{INSERT}, \texttt{UPDATE}, and \texttt{DELETE} requests operate only on the insert buffer. The key-value pair (or key-delete marker for deletes) is appended using two atomic counters: one atomic increment takes place before the write, and the other takes place after the data write, allowing concurrent writers. We ensure that concurrent lookups read values before the corresponding key to guarantee correct mappings. 

When a traverser writes into the last slot of the insert buffer, it sends a \emph{patch} request to the host via DMA. The host can then perform the required tree operations and propagate the tree updates via \emph{stitcher} threads back to the NIC-side tree~(see \Cref{subsecsec:update-cycle}). During this period, the NIC-side still has read-only access to the corresponding insert buffer. If a traverser cannot append to an insert buffer because it is full, the traverser re-enqueues for later processing. 

\texttt{GET} requests scan the insert buffer for recently updated or inserted values with the target key. Each leaf node has a separate insert buffer, which is unlikely that it is cached by the DPA thread. Hence, we issue a prefetch for the insert buffer while computing the model prediction for the default case. If the insert buffer contains the requested KV pair, an early exit occurs.
Otherwise, the array of leaf keys has to be scanned around the predicted position by the traverser thread. The leaf keys array resides in host memory. To reduce the DMA latency, we additionally prefetch the keys array during the model computation. Once the key position in the array is found, the value is read from the host memory via DMA.

\texttt{RANGE} requests reach the leaf node with the target key \(k_{\min}\) similar to \texttt{GET} requests. Then, the insert buffer is scanned, and all values within the requested range are added to a temporary location, which later becomes the response. The actual scan of the DMA arrays follows. When the scan reaches the end of a leaf, the traverser resumes at the next leaf by re-descending with the smallest key strictly larger than the last key returned.

Once the DPA thread finishes the request, it sends a response packet to the client. Response packets mirror the request layout but extend the type field with a status code and populate the value or range fields. For range responses, each packet carries at most 64 KV pairs to stay within a 1500-byte maximum transmission unit (MTU).

Tree traversal by DPAs is always lock-free because the tree structure is only changed on the host side, with changes being \emph{stitched} back to the NIC via RCU semantics~(see \Cref{subsecsec:update-cycle}). Therefore, DPA threads do not stall during tree traversal, and the tree maintains consistency at all times. 

\subsubsection{Learned index parameters}
We employ piecewise linear approximations (PLAs) with slope $a$, intercept $b$, and prediction $p(k) = a \cdot k + b$. For inner nodes, each segment contains a separate model, predicting the pivot for the node of the next level. For leaf nodes, the model predicts the location of the value. After prediction, the linear search visits at most $2\varepsilon$ keys in $[p-\varepsilon, p+\varepsilon]$. We enforce an error bound of $\varepsilon_{\text{inner}}=4$ for inner nodes (resulting in at most two cache line accesses) and $\varepsilon_{\text{leaf}}=8$ for leaf nodes (resulting in at most three DMA-accessed cache lines).

We use the greedy algorithms provided by \textsc{PGM}~\cite{FerraginaV20} to compute model parameters from sorted keys. The selected $\varepsilon$~values are fixed and enforced during training. 
To prevent memory segmentation and reduce allocation overheads during updates, we restrict the size of inner node segments and leaf nodes to hold at most 128 pivot keys and child pointers.

BlueField-3 DPAs do not offer floating-point operations~\cite{bluefield3-dpa}. Hence, all arithmetic operations use fixed-point numbers for slopes and intercepts. To keep precise calculations for the full 64-bit key space, we temporarily expand it to 128-bit during calculations (see also~\cite{ZhuSWCYYS25}).

\subsubsection{Hot entry caching on the NIC}
\label{sec:hotentry_cache}

\begin{figure}[t]
\includegraphics[width=\linewidth]{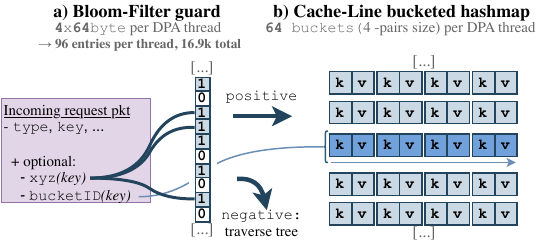}
\caption{DPA-side hot-entry cache.}
\label{fig:hotcache}
\end{figure}

Accesses to KV stores often follow a skewed distribution~\cite{CooperSTRS10, AtikogluXFJP12, CaoDVD20, YangYR21}, so caching hot entries can save tree traversals and DMA fetches. Therefore, each DPA traverser thread maintains a cache (see Figure~\ref{fig:hotcache}) consisting of (i) a three-way Bloom filter and (ii) a hash table composed of an array of buckets containing four KV pairs each. Each bucket is cache line sized for better memory utilization. 

To ensure that the same keys are always processed by the same DPA thread, clients send packets to a designated DPA thread by selecting its UDP port via a shared key hashing. The client also adds data required for cache lookups, such as the Bloom hash index and the hash bucket index, to reduce the effort required for DPA computations. Each Bloom filter contains 256 bits, which is the maximum size that fits into the remaining cache line of the traverser's thread context. The same cache line also contains important metadata for controlling RX/TX queues and is therefore unlikely to be evicted from the thread's L1 cache. Thus, the Bloom filter incurs no additional memory access cost.

We set the cache capacity to 96 entries per traverser thread, leading to an average false-positive rate of \(31\%\). With 176 traverser threads, a total of 16,896 entries can be cached. Assuming a dataset of 200M entries and an access pattern following a Zipfian distribution for $\alpha=1$, these cached entries make up more than 50\,\% of all requests. Randomly selecting which values to be cached yields an overall hit ratio of \(25\%\). We decided against actively tracking access patterns, as it requires additional counters and memory accesses. To guarantee consistency, the cache includes both keys and values to detect collisions of the key hash, and \texttt{UPDATE} and \texttt{DELETE} requests invalidate the corresponding cache entries.

\subsubsection{Maximizing delivery rate}
\label{par:maximizing_delivery_rate}

DPA-Store utilizes UDP for request and response transport. We chose not to use more robust protocols such as the Transmission Control Protocol (TCP) because the DPA invocation is event-based and therefore unsuitable to implement a TCP stack that requires session windows, transmission buffers, sequence tracking, and time-based re-sends. Instead, DPA-Store handles buffer overflows on the server by monitoring response rates at the client side. If the client does not receive a response in a timely fashion, it re-sends the request. This problem is common for network-intercepting designs, e.g., for \textsc{KV-Direct} and \textsc{Mica}~\cite{LiRXLXPCZ17,LimHAK14}.

The per-thread DPA receive queues (256 packets for each traverser) allow for a maximum number of 45,056 in-flight packets. This enables a DPA-Store instance to handle many clients. Still, DPA-side overflows can occur when the key hashing scheme, which assigns packets to DPA cores, causes extreme skewness, overwhelming single DPA threads. Therefore, if the response rate on a client exceeds a threshold, the clients can gradually send \texttt{GET} requests to different traverser threads using an alternative hashing scheme. To prevent hot-entry cache incoherency at this point, the client removes the hash metadata from the request, bypassing the cache.

\subsection{Update Cycle}
\label{subsecsec:update-cycle}

This subsection explains how buffered writes become structural updates without blocking traverser threads. We introduce concurrent \emph{patch} threads on the host and \emph{stitch} threads in the DPA subsystem. DPA-Store guarantees that the NIC-side tree is consistent at all times by employing RCU semantics, so nodes are never modified in-place, but exchanged via atomic pointer swaps. More complex tree updates are executed bottom-up until reaching the last node to be modified. Its address is updated in its parent via a pointer swap. This stitching procedure is transparent to all traverser threads. 

The cache consistency protocols of the DPA subsystem guarantee that a node's address is always consistent over all DPA threads~\cite{bluefield3-dpa}. However, a traverser may descend an outdated sub-tree. To maintain overall consistency, outdated nodes and host-side DMA locations are only garbage collected after every traverser has moved on to its next request using epochs.

\subsubsection{Host-side patching}

The DPA traverser threads append insert, delete, and update operations to the per-leaf insert buffers. When an insert buffer becomes full, the corresponding traverser enqueues a patch request to its dedicated host-memory queue via DMA. Only one patch request can be triggered per insert buffer since only the traverser that fills the buffer is allowed to emit the patch request. On the host, a number of patcher threads (e.g., four) process the patch requests and update the host-side tree. If the patch request contains only \texttt{UPDATE} operations, the patcher modifies the values accordingly. In this case, the patcher thread notifies the DPA-side stitcher threads to clear the insert buffer and perform no further action. 

If the patch contains \texttt{INSERT} or \texttt{DELETE} operations, the patcher first merges the key-value pairs with the existing leaf contents into a temporary, sorted array. 
The resulting array is then partitioned using PLA segmentation for $\varepsilon_{\text{leaf}}$ with each segment becoming a new leaf node. This results in either a single segment that fits within the leaf capacity or multiple segments. When splitting becomes necessary, we limit segment sizes by a \emph{retrain bound} (0.25$\times$capacity). The resulting leaf nodes are sparsely populated, and future patch requests may be absorbed without another split, reducing the overall number of split operations.

\begin{figure}[t]
\includegraphics[width=\linewidth]{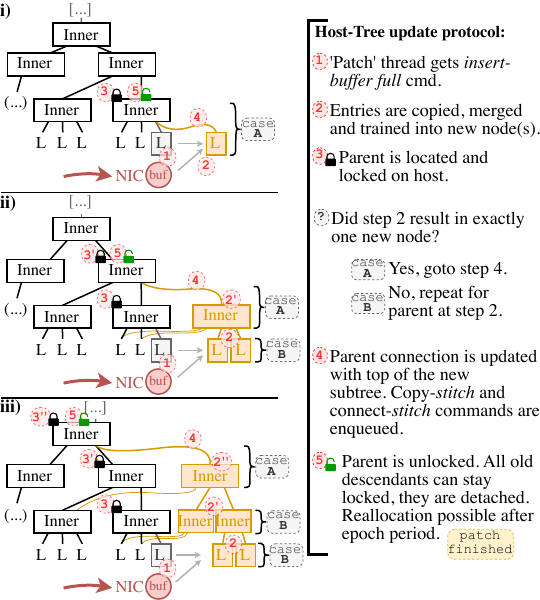}
\caption{Host-side \emph{patching} protocol.}
\label{fig:insert-protocol}
\end{figure}

After the leaf nodes are updated on the host, the patcher thread locates the parent of the original leaf by descending from the root. It then locks the parent for exclusive access, making simultaneous updates originating from sibling nodes wait for the current update to finish. We avoid explicit parent pointers because maintaining bidirectional references under concurrency requires complex synchronization schemes.

If the retraining produced exactly one leaf, a single pointer swap in the parent is sufficient. Otherwise, the parent must also be rebuilt. We merge the new child pivots with the parent entries and apply PLA segmentation with $\varepsilon_{\text{inner}}$. PLA ranges in this step represent the inner-node segments. If the segmentation yields more ranges than the maximum allowed per node, multiple inner nodes are created, each containing a balanced number of segments. Similar to splitting leaf nodes, we limit the number of segments within inner nodes using the \emph{retrain bound}.
Overall, this process is iterated bottom-up for every level of the tree, only stopping if the parent node does not need to be split or the root node has been reached.

Once the host tree contains the new nodes, we propagate the changes to the NIC-side tree using our stitcher threads. We enqueue the individual node updates as \texttt{COPY} stitches to the stitcher queues. Afterward, a single \texttt{CONNECT} stitch command is issued that translates to a single pointer swap, making previous \texttt{COPY} stitches effective (see Figure~\ref{fig:insert-protocol}).

\begin{figure}[t]
\includegraphics[width=\linewidth]{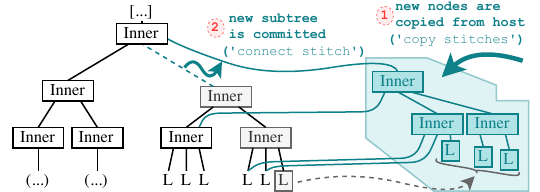}
\caption{The NIC-tree stays valid while \emph{stitches} are applied.}
\label{fig:stitching-validity}
\end{figure}

\subsubsection{NIC-side stitching}
\label{sec:nic-side-stitching}

\begin{figure}[t]
\includegraphics[width=\linewidth]{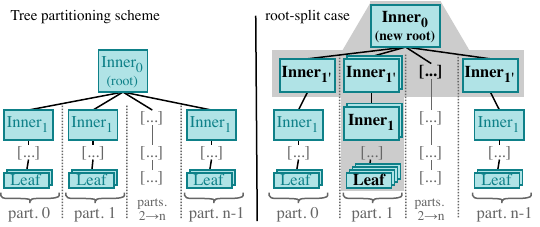}
\caption{Tree partitioning to allow concurrent \emph{stitches}.}
\label{fig:stitch-protocol}
\end{figure}

The NIC-side DPA stitcher threads apply stitch commands submitted through stitcher thread queues. Threads process incoming \texttt{COPY} and \texttt{CONNECT} stitches in order, redoing the host-tree changes on the NIC-side~(\Cref{fig:stitching-validity}). The host has pre-calculated every destination address in the DPA memory space, and the stitcher threads do not have to allocate any memory. Instead, they execute the provided commands on the pointers passed from the host.
To enable concurrent stitching, we partition the tree beneath the first level of inner nodes (Figure~\ref{fig:stitch-protocol}, left). Parallel stitcher threads (four by default) work on a dedicated partition of the tree. The host is aware of the mapping of stitcher threads to partitions and assigns stitch commands accordingly. 

Ordering of stitches and single-thread ownership eliminate most race conditions. However, root-level updates can lead to colliding updates: (i) a \texttt{CONNECT} could point to a leaf's memory address that was not previously created using a \texttt{COPY} stitch, or (ii) newly copied and connected nodes could reference nodes underneath that are not yet available to the NIC-side tree.
Therefore, we have implemented two safeguards. First, we delay the execution of stitches targeting nodes originating directly from a root split until their destination is available. We do this by using UIDs to probe whether the target node of a \texttt{CONNECT} stitch is already in place. Second, the stitch that installs an updated root node is blocked by a queue fence until earlier updates have completed. To preserve the partitioning scheme, the host ensures that root splits maintain balanced partitions and distribute new top-level nodes across all partitions (Figure~\ref{fig:stitch-protocol}, right-hand side).

\subsubsection{Memory reclamation}

NIC-side sub-trees become obsolete after a new version is installed using \texttt{COPY} and \texttt{CONNECT} stitches, in which case the leaf memory must be eventually freed. However, traverser threads may still descend into sub-trees containing the affected nodes. Therefore, we perform epoch-based reclamation of obsolete nodes, using all DPA threads' incoming and outgoing packet counters to compute a global epoch value.

\subsubsection{Bulk loading}

Bulk loading partitions a set of sorted KV pairs into PLA segments for the configured leaf error bound $\varepsilon_{\text{leaf}}$ inside the host. Each PLA segment becomes a leaf node. Once all leaf nodes are available, their leading keys form the scaffolding for building upper levels. We apply the same PLA construction bottom-up using the first keys of child nodes for the nodes of the next-higher level, enforcing an error bound of $\varepsilon_{\text{inner}}$ to regulate tree fan-out. This recursive process continues until a single inner node remains, which becomes the root node.

Throughout bulk loading, the host enqueues \texttt{COPY} and \texttt{CONNECT} stitch commands to the stitcher queues. The stitcher threads assemble the initial tree following the enqueued host commands, ending with a final root-pointer stitch that makes the structure visible to traverser threads. As the host determines the tree partitions by selecting a stitcher queue for every tree update, it implicitly creates the partitions, and no further actions on the NIC-side are necessary.

\section{Evaluation}

In this section, we first provide an overview of the experimental setup, followed by an in-depth analysis of DPA-Store. Finally, we compare DPA-Store with ROLEX~\cite{Li0ZCS23}, a state-of-the-art RDMA-based KV store.

\subsection{Experimental Setup}
Unless stated otherwise, our test environment consisted of one server and six client machines connected through a 100\,Gb/s Dell PowerSwitch S5232F. The switch is Ethernet-based and supports RDMA over Converged Ethernet (RoCE). The clients use dual-socket AMD Epyc 7301 CPUs (32 cores) with Mellanox ConnectX-5 NICs. Clients leverage DPDK~\cite{dpdk} to send requests to circumvent OS bottlenecks. The server uses a BlueField-3 B3140L on a system with a 32-core AMD Epyc 9354P and 128\,GB of DDR5 memory. In all experiments, including DPA-Store, the BlueField-3 operates in NIC mode, i.e., with DPA DMA access to the host and disabled ARM cores. When the DPA subsystem is not in use, the BlueField-3 behaves like a ConnectX-7 NIC. This allows us to evaluate related work on a conventional host/NIC setup. At the same time, DPA-Store executes on the same hardware and software stack, ensuring a fair comparison. 

Many learned indexes are sensitive to key distributions. 
We ensure rigorous testing of our learned index by using the common SOSD datasets~\cite{MarcusKRSMK0K20, BellomoCRFO25}. They consist of the synthetic \emph{sparse} and \emph{dense} datasets, as well as real-world datasets derived from Facebook (\emph{face}), Amazon (\emph{amzn}), Wikipedia (\emph{wiki}) and OpenStreetMap (\emph{osmc}) workloads. The \emph{sparse} dataset randomly selects keys from the full 64-bit range, while our \emph{dense4x} selects 50M keys from a consecutive range of 200M keys. Unless stated otherwise, we bulk load 25M entries before each experiment and use a value of $\alpha=0.99$ for skewed key popularities according to Zipf's Law. Similar to most related work, we use 64 bit values besides our 64 bit keys.

We averaged throughput values over four runs for long-running benchmarks (e.g., \texttt{GET}) and eight runs for short-running ones (e.g., inserts of new keys). Latency measurements were gathered from all clients and combined. Client start times were synchronized via MPI. Measurements showed a standard deviation of less than 5\% for throughput and less than 9\% for latencies.

\subsection{DPA-Store Evaluation}

In this section, we investigate the memory consumption, the effects of varying the learned index, the parameters of the client, and thread counts for DPA and host threads on throughput and latency. Using our findings, we deduce default values for DPA-Store. Furthermore, we analyze whether different BlueField-3 models or changes to the BlueField-3 hardware influence performance.

\subsubsection{Memory consumption}

We evaluated the overheads of the index structure compared to the raw KV data on the host-side for 50M inserted KV pairs (\Cref{tab:memory_consumption}). Because \emph{face} and \emph{osmc} show the highest memory consumptions and overheads for values of $\varepsilon_{\text{inner}}=4$ and $\varepsilon_{\text{leaf}}=8$, we demonstrate that their memory impact can be significantly reduced with $\varepsilon = 16$. In the following benchmarks, we select the larger values for $\varepsilon$ for those datasets.

Compared to other learned indexes, DPA-Store chooses small $\varepsilon$ values to minimize the costs for finding the key in the area around the predicted position on latency-constrained DPA memory. For example, \textsc{ROLEX} uses $\varepsilon \in \{128, 256\}$, allowing it to maintain much larger nodes and thus reducing per-node memory overheads. \textsc{ROLEX} reports a metadata overhead of 6.5\,\%
for cache data for a 500\,M dataset, assuming 16\,B KV pairs~\cite{Li0ZCS23}. However, it requires all clients to hold a separate cache, increasing memory overhead proportional to the number of clients in addition to the number of keys.

\begin{table}[t]%
    \vspace{-0.8em}\caption{Relative overhead and NIC-side memory consumption for 64-bit key distributions and 50M entries. }\vspace{2mm} 
    \centering%
    \begin{tabular}{lcc}
        \makecell{\phantom{x}\\\textbf{Dataset (50M)}} & \makecell{\textbf{Rel. overhead}\\\textbf{of index}} & \makecell{\textbf{Overall DPA}\\\textbf{memory req.}} \\
        \hline
        sparse & 32\% & 207 MB\\
        dense4x & 26\% & 169 MB\\
        wiki & 23\% & 147 MB\\
        amzn & 54\% & 346 MB\\
        osmc & 74\% & 472 MB\\
        face & 104\% & 672 MB\\
        osmc ($\varepsilon_{\text{in}}=\varepsilon_{\text{lf}}=16$) & 35\% & 228 MB\\
        face ($\varepsilon_{\text{in}}=\varepsilon_{\text{lf}}=16$) & 52\% & 332 MB\\
    \end{tabular}
    \label{tab:memory_consumption}
\end{table}

\subsubsection{DPA and host thread counts}

\begin{figure}[t]
\includegraphics[width=\linewidth]{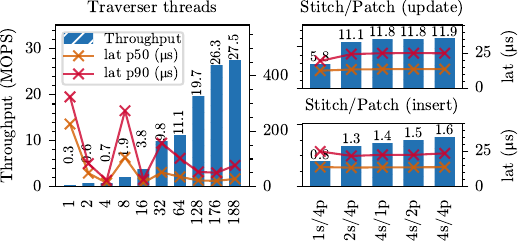}
\caption{DPA-store throughput and latencies for different numbers of traverser, patcher, and stitcher threads. The left plot shows the number of traverser threads for \texttt{GET} requests. The right plot varies patcher/stitcher thread count for \texttt{INSERT}- and \texttt{UPDATE}-only workloads with 176 traversers.}
\label{fig:traverser_thread_scaling}
\end{figure}

In this subsection, we evaluate the impact of a varying number of traverser, patcher, and stitcher threads on the DPA-Store performance. \Cref{fig:traverser_thread_scaling} (left) shows the throughput and latencies of a \texttt{GET}-only workload with uniform key popularity on the sparse dataset. We observe that the throughput scales proportionally to the number of traverser threads. Latencies exhibit higher variance with few traverser threads but stabilize with a larger number of threads. This is because the few threads are overwhelmed by the number of incoming packets, causing them to fail to process requests in a timely fashion. 
For values larger than 16, multiple DPA hardware cores are used, resulting in lower latencies.

Throughput increases slightly beyond 176 threads, which equals eleven hardware DPA cores. The remaining 13 DPA threads then reside on the last physical core. We found that using the last available hardware core for both traverser and stitcher threads leads to worse throughput for \texttt{INSERT} operations. This is a result of the hardware scheduler that prioritizes NIC doorbell events. These events invoke traverser threads and therefore stall the execution of stitchers. We observed a 14\% degradation in throughput for \texttt{INSERT} when mixing both types of DPA threads on one core. For both \texttt{INSERT} and \texttt{UPDATE} workloads, throughput flattens for more than four patcher and stitcher threads. We note that the number of host-side patcher threads is limited in throughput by the NIC-side stitcher threads (see Section~\ref{sec:insert_update_analysis} for a detailed analysis).

\textbf{Lessons learned:} Prioritizing NIC doorbell events by the BlueField-3 requires to partition the 189 available DPA threads into 176 traverser threads on 11 physical cores and four stitcher threads on a dedicated core. Furthermore, we configure DPA-Store to run four host-side patcher threads.

\subsubsection{Client-side queue depth}

\begin{figure}[t]
\includegraphics[width=\linewidth]{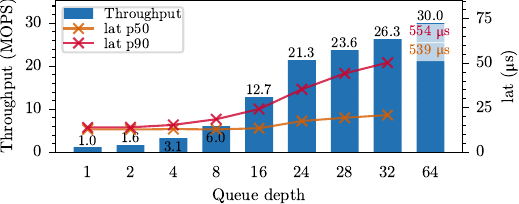}
\caption{DPA-Store \texttt{GET} throughput and latency using different client-side queue depths.}
\label{fig:queue_depth_scaling}
\end{figure}

In the following, we investigate the impact of the client-side queue depth and the resulting total number of in-flight requests on throughput and latencies. Each of the 6 client nodes are running 31 threads. Threads issue between 1 and 64 concurrent requests. NIC-side hash tables ensure that outgoing and corresponding incoming packets are handled by the same thread and that requests lacking an acknowledgment are resent after a timeout. Figure~\ref{fig:queue_depth_scaling} shows how client-side queue depth affects \texttt{GET} performance under a uniform key distribution.

Throughput increases significantly up to a client queue depth of 32 with a maximum of $5,952$ in-flight requests. After this point, throughput continues to rise; however, latency increases beyond acceptable levels because the processing rate of DPA-Store cannot keep pace with the request submission rate. Accordingly, a queue depth of 32 is used for \texttt{GET} requests throughout all subsequent experiments. We similarly set a queue depth of 18 for insert and range workloads. Integrating an adaptive flow control is left as future work.

\begin{figure}[t]
\includegraphics[width=\linewidth]{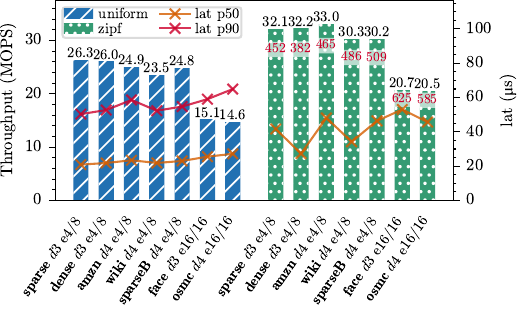}
\caption{Results of DPA-Store using \texttt{GET}-only workloads.}
\label{fig:dpastore_get_performance}
\end{figure}

\subsubsection{Effect of tree depth and index error bound}

We evaluate the effect of tree depth and the error bound $\varepsilon$ on the throughput and latency of DPA-Store. To demonstrate the effect of the tree depth, we include the \emph{sparseBig} dataset in this section. Bulk loading \emph{sparseBig} (consisting of 50M KV pairs) results in a tree depth of four, whereas \emph{sparse} requires three tree levels. Figure~\ref{fig:dpastore_get_performance} shows \texttt{GET} performance for different datasets under uniform and skewed key popularity. Under uniform access, we observe a slight reduction in throughput for deeper trees because of the extra accesses to an additional inner node. The more significant performance degradation when choosing a larger $\varepsilon$ value for the datasets \emph{face} and \emph{osmc} illustrates the impact of additional cache-line accesses required when verifying the model predictions (see also DPA-Store performance model in Section \ref{sec:memory_bandwidth_optimizations}).  

With skewed key popularity, the hot-entry cache increases throughput by up to 30\,\%, which meets our expectations in Section~\ref{sec:hotentry_cache}. However, tail latencies increase because some traverse threads receive a disproportionate share of requests, leading to longer wait times in the DPA queues.  

\subsubsection{\bptree\ comparison}

We chose a learned index as the fundamental data structure behind DPA-Store. To motivate this decision, we compare the throughput and latencies for \texttt{GET} operations of the learned index tree against a \bptree\ directly after bulk-loading the data. For the \bptree, we therefore set $\varepsilon=\infty$ when building the tree nodes, leading to fully packed 2 kByte nodes with 128 entries each. For lookups inside nodes, the \bptree\ uses binary search. We use 176 traverser threads for both variants.

\begin{figure}[b]
\includegraphics[width=\linewidth]{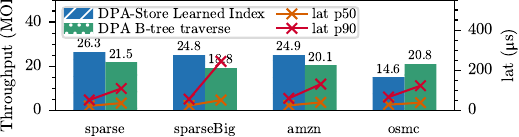}
\caption{Comparing \bptree\ and learned DPA traversals.}
\label{fig:btree_comparison}
\end{figure}

Figure~\ref{fig:btree_comparison} shows that latencies are mostly higher for the \bptree\ than for the learned index. With the default $\varepsilon$, the learned index achieves higher throughput on the \emph{sparse}, \emph{sparseBig}, and \emph{amzn} datasets, even though the more densely populated nodes in the \bptree\ yield lower tree depths.
The \emph{osmc} dataset is configured to use $\varepsilon=16$, resulting in better throughput for the \bptree. This shows that optimized learned traversals of DPA-Store depend on small values for $\varepsilon$ to be effective.

\subsubsection{Analysis of \texttt{GET} operations and memory accesses}
\label{sec:memory_bandwidth_optimizations}

The previous two subsections have shown that DPA-Store depends heavily on the performance of the memory subsystem, particularly the access latency of the DPA-addressable DDR5 memory. Prior work~\cite{Chen0FSMQZSZLW24} reports average memory access times of 465\,ns for DPA memory, compared with 910\,ns for DMA accesses to host memory. In the following, we model the minimal duration of a full tree traversal, assuming our default parameters of $\varepsilon_{\text{inner}}=4$ and $\varepsilon_{\text{leaf}}=8$.

The first cache line accessed in every inner node includes the node's metadata and the segment's first keys. The second cache line contains each segment's model. The next one or two cache lines (depending on the node fullness) contain the pivots, and the fifth cache line contains the child pointer. Assuming nodes to be filled 50\% on average, this results in an average of 4.5 cache lines per inner node. Similarly, leaf nodes require one cache line for metadata and up to three DMA operations to fetch the key and one for the value. Note that we assume insert buffers to be empty in this analysis, requiring no additional memory accesses.

For a tree of depth 3, we access two inner nodes and one leaf node, resulting in overall memory access times of 
\begin{align*}
2 \cdot \delta_{inner} + \delta_{leaf} & = 2 \cdot 4.5 \cdot 0.465\mu s + 0.465 \mu s + 2\cdot 0.91 \mu s \\
 & = 6.47\mu s
\end{align*}
Note that the cache lines for the leaf key are sequential on the host and collapse into a single DMA. Assuming DPA scheduling overlaps one thread’s computation with another thread’s memory accesses, the maximal throughput is $176 /6.47\mu s = 27.2$\,MOPS, provided none of the relevant cache lines are cached.

If we further assume the root node's first cache lines (node metadata, segment keys, and models) are cached for all DPA threads, and the two memory accesses are replaced by L3 access times of $64\,ns$ each, we compute $\delta_{\text{root}}=1.2905$ and obtain a maximal throughput of $31.05$\,MOPS. While this model is an approximation, our evaluation of DPA-Store achieves similar results (e.g., \Cref{fig:traverser_thread_scaling}).

As detailed in \Cref{subsecsec:nicside-request-path}, we optimized the traverser path to prefetch cache lines optimistically, overlapping memory access with compute. Due to the relatively weak per-thread compute capabilities, we measured an improvement of 19\,\% for this optimization.

\textbf{Lessons learned:} \texttt{GET} performance is limited by the access latencies to DPA memory and could increase significantly if memory access latencies would be comparable with latencies of standard CPUs. In case of memory latencies of 100\,ns, \texttt{GET} latencies could decrease to less than 2.82 $\mu$s and throughput could increase to more than 62 MOPS. Decreased memory latencies would also allow to increase $\varepsilon$ values and to become more memory efficient.

\subsubsection{Bulk load performance}
\label{sec:bulk_load_performance}

We evaluated the bulk load throughput using a 50M \emph{sparse} dataset and four stitcher threads. Traverser threads are disabled during bulk loading. The full bulk load is finished on the host after an average runtime of 1,643\,ms. The copying and processing of stitch requests accumulates to 1,605\,ms. The bulk load copies 192 MByte of tree data into DPA memory, resulting in a host-to-DPA memory bandwidth of only 120 MByte/s. The reason for this low bandwidth is that it is not possible to efficiently write fine grained data that targets various addresses directly from host to the DPA memory. Instead, the stitcher threads have to poll for incoming stitches per DMA and then to first load the data into the DPA thread and then into DPA memory. 

\textbf{Lessons learned:} Bulk load performance and, as shown in the following section, insert performance could be greatly increased if the BlueField-3 would support efficient transfers from host memory to DPA memory.

\subsubsection{\texttt{INSERT}/\texttt{UPDATE} performance}
\label{sec:insert_update_analysis}

\begin{figure}[b]
\includegraphics[width=\linewidth]{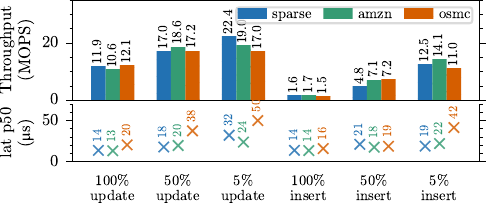}
\caption{DPA-Store performance \texttt{UPDATE}s and \texttt{INSERT}s.}
\label{fig:insert_update_performance}
\end{figure}

We evaluate throughput and latency for \texttt{INSERT} and \texttt{UPDATE} operations, mixed with varying degrees of \texttt{GET} operations, using the \emph{sparse}, \emph{amzn}, and \emph{osmc} datasets (see \Cref{fig:insert_update_performance}). 

Stitch requests for \texttt{UPDATE}-only patches require no data to be copied to DPA memory. Instead, only the host-side values are updated, and a stitch command resets the original DPA leaf's insert buffer. Here we reach up to 12.1\,MOPS.
We particularly observe low throughput for \texttt{INSERT} operations across all three datasets with at most 1.7\,MOPS.
As soon as nodes are retrained on the host-side from the leaf upward,  \texttt{COPY} stitches are transferred. For leaves, only model parameters and DMA addresses are transferred, while inner nodes copy their complete pivots and child pointers.
As we saw already for the bulk load performance, which uses the same stitching method as runtime updates, we are heavily constrained by the BlueField-3's inability to efficiently write data from host to device. The bandwidth of data that is stitched during runtime in an \texttt{INSERT}-heavy scenario cannot exceed the low rate we measured for bulk loading. Although the polling stitchers perform below our expectations, we see potential generations of SmartNICs will address this issue and instead focus on maximizing the \texttt{GET} and \texttt{RANGE} performance.

\subsubsection{BlueField-3 model comparison}

\begin{figure}[b]
\includegraphics[width=\linewidth]{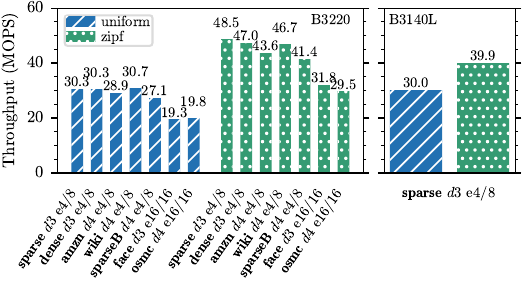}
\caption{\texttt{GET} throughput for different BlueField-3 models.}
\label{fig:dpastore_get_performance_max}
\end{figure}

\begin{figure*}[t]
\centering
\includegraphics[width=0.96\linewidth]{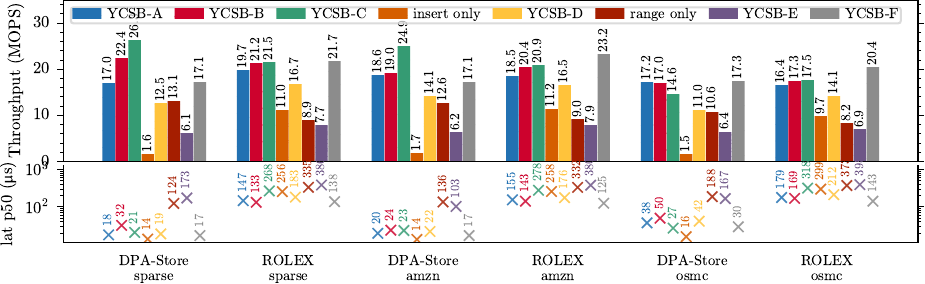}
\caption{YCSB workloads on \emph{sparse}, \emph{amzn}, and \emph{osmc} datasets for DPA-Store and \textsc{ROLEX}.}
\label{fig:ycsb_comparison}
\end{figure*}

We run tests on an additional setup to evaluate the B3220 model of the BlueField-3 family. Unlike the B3140L with single-channel memory and one network port, the B3220 features dual-channel DPA memory and two network ports. The number and type of DPA cores, however, remain the same. Unfortunately, comparable client nodes were unavailable and we had to generate requests with another BlueField-3 B3220 NIC via DPA programs. This allowed us to saturate the throughput of DPA-Store but the client card lacked compute performance to evaluate response latencies. We also disabled the restriction on the number of in-flight requests for the B3140L. Therefore, the throughput of DPA-Store has been saturated for both models, enabling a fair comparison.

Figure~\ref{fig:dpastore_get_performance_max} shows \texttt{GET}-only workloads on different datasets as well as uniform and skewed key popularities for both NICs. We observe that throughput for the \emph{sparse} dataset with uniform key distributions is the same for both NICs. This shows that \texttt{GET} throughput is dominated by DPA memory latencies and that moving from single- to dual-channel memory has no direct effect on DPA-Store. DPA-Store's \texttt{INSERT}, \texttt{UPDATE}, and \texttt{RANGE} throughput also do not differ significantly. 

However, with a skewed key distribution for \texttt{GET}-only workloads, the B3220 shows higher throughput than the B3140L, reaching 48.5\,MOPS, whereas the B3140L reaches 39.9\,MOPS. Additionally, we ran ping tests that return packets by the receiving DPA thread without memory access. For ping, the B3140L model reached 44.9\,MOPS, whereas the B3220 delivered 69\,\% more throughput. Because cached requests trigger only a single memory access and uniform accesses remain unchanged, these results demonstrate that the dual-port B3220 features stronger packet-matching hardware and therefore can process small requests faster.

\subsection{Comparison with ROLEX}

We compared DPA-Store against \textsc{ROLEX}~\cite{Li0ZCS23}, the fastest RDMA KV store able to run on modern hardware and being available, to highlight the effects of different architectural choices. We evaluated \textsc{ROLEX} in the same configuration in which we evaluated DPA-Store. This led to better throughput results for \textsc{ROLEX} compared to its own published results but also increased its latencies. 

We evaluated both solutions using YCSB workloads~\cite{CooperSTRS10} consisting of six scenarios: A (50\,\% reads, 50\,\% updates), B (95\,\% reads, 5\,\% updates), C (100\,\% reads), D (95\,\% reads, 5\,\% inserts), E (95\,\% range, 5\,\% inserts), and F (50\,\% reads, 50\,\% read-modify-update). Additionally, we include experiments for 100\,\% inserts and 100\,\% range queries covering 10 adjacent keys.
The experiments are executed using \emph{sparse}, \emph{amzn}, and \emph{osmc} datasets, which resulted in different tree depths and error bound configurations for DPA-Store (cf. \Cref{fig:dpastore_get_performance}) and we measured with uniform key distributions, reducing the effect of the hot-entry cache.

Figure~\ref{fig:ycsb_comparison} shows that DPA-Store exceeds \textsc{ROLEX} throughput for the \emph{amzn} and \emph{osmc} datasets for YCSB-A and for all \texttt{RANGE}-only workloads. For \texttt{GET}-only workloads, DPA-Store achieves higher throughput and lower latencies compared to \textsc{ROLEX} on \emph{sparse} and \emph{amzn}. However, \textsc{ROLEX} achieves better results on \emph{osmc} due to its larger values of $\varepsilon$ being more suitable for this dataset.

ROLEX shows higher throughput for \texttt{INSERT} workloads. Its clients are capable of directly RDMA'ing \texttt{INSERT} requests into the server memory, only being restricted by the server's memory performance. Due to its stateful clients, \textsc{ROLEX} can decouple its model re-training lazily, propagating the load of tree updates to the client side. This approach distributes the load more evenly. However, it requires in-depth client logic and state-keeping. On the other hand, DPA-Store imposes no state on the client, expecting only key hashing for thread allocation to take place. The \texttt{INSERT} workload is shown only for comparison. In real applications values would be bulk loaded before a run, and YCSB workloads represent typical usage with at most 5\% inserts. 

In workloads where inserts of new keys constitute only a small fraction compared to \texttt{GET} or \texttt{RANGE} requests, such as YCSB-D and YCSB-E, DPA-Store nearly reaches \textsc{ROLEX} performance. Updating existing keys is not restricted as much by the host-to-DPA bottleneck because raw updates do not change inner nodes, and no additional copy stitches are sent. Consequently, even for high update ratios, as in YCSB-A, DPA-Store can surpass \textsc{ROLEX} for \emph{amzn} and \emph{osmc}. 

\Cref{fig:ycsb_comparison} shows lower latencies for DPA-Store compared to \textsc{ROLEX} in all experiments. While DPA-Store may require passing the relatively slow PCIe bus to the host for DMA access when performing a \texttt{GET} operation, the hot cache and insert buffer reduce the number of overall DMA operations. ROLEX requires every operation to execute RDMA operations, for uncached cases even multiple roundtrips, causing noticeable contention delays for more in-flight requests.

\textbf{Lessons learned:} DPA-Store is, at the cost of a more complex NIC hardware, faster for most  workloads than the state-of-the-art ROLEX KV store but suffers from lower \texttt{INSERT} throughput. However, DPA-Store could significantly surpass ROLEX's throughput for all workloads, as shown in previous sections, if the next Bluefield generation includes a better DPA memory interface and the ability to efficiently transfer data from host to DPA memory.

\section{Conclusion}
We have proposed DPA-Store, a KV-store residing on a SmartNIC that removes OS latencies from all operations. By carefully analyzing the internal hardware of the BlueField-3, we have designed DPA-Store to maximize operation throughput. We have offloaded the computation-heavy learned index maintenance to the host while keeping the tree traversal lock-free and highly concurrent. We have shown that DPA-Store is competitive with the ROLEX and that small changes to the Bluefield-3 could help to significantly surpass these results.

\section*{Acknowledgement}
This work has been supported by the project \emph{Big Data in Atmospheric Physics (BINARY)}, funded by the Carl Zeiss Foundation
(P2018-02-003) and the project \emph{ScalNEXT:\@ Optimierung des Datenmanagements und des Kontrollflusses von Rechenknoten für Supercomputing} (16ME0688).
\\
Further, this research was conducted using hardware of the Institute for Automation of Complex Power Systems at RWTH Aachen University.

\bibliographystyle{plainurl}
\bibliography{bib}

\end{document}